\documentclass[aip,jap,reprint]{revtex4-1}%
\usepackage{graphicx}
\usepackage{dcolumn}
\usepackage{bm}
\usepackage{amsmath}%
\usepackage{amsfonts}%
\usepackage{amssymb}

\begin{document}
\title
{Monte Carlo study of carrier-light coupling in terahertz quantum cascade lasers}
\author{Christian Jirauschek}
\email{jirauschek@tum.de}
\homepage{http://www.nano.ei.tum.de/noether}
\affiliation
{Emmy Noether Research Group ``Modeling of Quantum Cascade Devices``, Institute for Nanoelectronics, Technische
Universit\"{a}t M\"{u}nchen, D-80333 Munich, Germany}
\date{\today, published as Appl. Phys. Lett. 96, 011103 (2010)}
\begin{abstract}
We present a method for self-consistently including the optical cavity field into Monte Carlo-based carrier transport simulations. This approach allows for an analysis of the actual lasing operation in quantum cascade lasers, considering effects such as gain saturation and longitudinal mode competition. Simulation results for a terahertz quantum cascade laser are found to be consistent with experiment.
\end{abstract}
\maketitle

The development of innovative types of quantum cascade lasers (QCLs) and
subsequent design optimization has gone hand in hand with detailed modelling,
involving more and more sophisticated simulation
tools.\cite{2001ApPhL..79.3920K,2001ApPhL..78.2902I,2004ApPhL..84..645C,2005JAP....97d3702B,2007JAP...101f3101G,2007JAP...101h6109J,2009JAP...105l3102J,2001PhRvL..87n6603I,2007JAP...102k3104N,2009PhRvB..79s5323K}
Especially in the terahertz regime, there is still plenty of room for
improvement of the structures, for example in terms of output power,
efficiency and temperature performance.\cite{2007NaPho...1..517W} In this
context, detailed carrier transport simulations have proven very useful,
accounting for inter- and intrasubband processes alike and yielding not only
level occupations, but also kinetic carrier distributions within each of the
levels. Well-established approaches include the semiclassical ensemble Monte
Carlo (EMC)
method,\cite{2001ApPhL..79.3920K,2001ApPhL..78.2902I,2004ApPhL..84..645C,2005JAP....97d3702B,2007JAP...101f3101G,2007JAP...101h6109J,2009JAP...105l3102J}
and quantum transport simulations based on the density
matrix\cite{2001PhRvL..87n6603I} or non-equilibrium Green's functions
formalism.\cite{2007JAP...102k3104N,2009PhRvB..79s5323K} Notably, these
simulations have up to now almost exclusively focused on the carrier
transport, completely neglecting the optical cavity field. An exception is a
Monte-Carlo-based study of the coupled cavity dynamics and electron transport,
however only considering carrier interaction with photons and
phonons.\cite{2005RPPh...68.2533I} Carrier transport simulations allow for an
analysis of the unsaturated optical gain, indicating if the investigated QCL
structure will lase at all under the assumed conditions. However, no
statements about the actual lasing operation, including the emitted optical
power, the electric current and the carrier distributions, can be made.
Evidently, an inclusion of the lasing action in the simulation would be
desirable for many practical reasons, above all the device optimization with
respect to the output power and wall-plug efficiency. Furthermore, such an
analysis could provide insight into the carrier dynamics on a microscopic
level, yielding an improved understanding of the light-matter interaction in
such devices. In the following, we introduce an approach which allows for a
straightforward implementation of the laser field into EMC simulations,
without a significant increase of the numerical load. The presented scheme
considers the coupled cavity field and carrier transport dynamics in a
completely self-consistent manner, accounting for effects such as gain
saturation and longitudinal mode competition. We present simulation results
for a terahertz QCL, which are found to be consistent with experimental data.

Induced optical transitions between an initial and a final level, in the
following denoted by $i$ and $j$, are commonly described in terms of a
spectral power gain coefficient $g_{ij}\left(  \omega\right)  $ and the
population change associated with the induced emission and absorption
events,\cite{2003noop.book.....B}%
\begin{subequations}%
\label{twol}%
\begin{align}
g_{ij}\left(  \omega\right)   &  =\frac{\omega_{ij}}{\left|  \omega
_{ij}\right|  }\frac{\pi\omega Z_{0}}{Vn_{0}\hbar}\left|  d_{ij}\right|
^{2}\left(  p_{i}-p_{j}\right)  \mathcal{L}_{ij}\left(  \omega\right)
,\label{twol_a}\\
\left.  \partial_{t}p_{i}\right|  _{\mathrm{ind}}  &  =I\frac{\pi Z_{0}}%
{n_{0}\hbar^{2}}\left|  d_{ij}\right|  ^{2}\left(  p_{i}-p_{j}\right)
\mathcal{L}_{ij}\left(  \omega\right)  . \label{twol_b}%
\end{align}
\end{subequations}%
Here, $g_{ij}\left(  \omega\right)  >0$ corresponds to gain, and
$g_{ij}\left(  \omega\right)  <0$\ indicates loss. The constants $Z_{0}$ and
$\hbar$ denote the impedance of free space and reduced Planck constant,
respectively. $V$ and $n_{0}$ are the volume and refractive index of the gain
medium. The line shape as a function of the angular optical frequency $\omega$
is given by
\begin{equation}
\mathcal{L}_{ij}\left(  \omega\right)  =\frac{1}{\pi}\frac{\gamma_{ij}}%
{\gamma_{ij}^{2}+\left(  \omega-\left|  \omega_{ij}\right|  \right)  ^{2}},
\label{L}%
\end{equation}
where $\gamma_{ij}$ and $\omega_{ij}=\left(  E_{i}-E_{j}\right)  /\hbar$
denote the optical linewidth and resonance frequency of the transition.
$E_{i,j}$ and $p_{i,j}$ are the level eigenenergies and occupations, and
$d_{ij}$ is the corresponding transition matrix element. Furthermore, $I$
denotes the optical intensity.

In a semiclassical EMC simulation, the carrier transport is modeled by a
stochastic evaluation of the inter- and intrasubband scattering events for a
large ensemble of discrete particles. Each carrier is at a given time
described by its quantum state $\left|  i_{n},\mathbf{k}_{n}\right\rangle $
(which then has an occupation $p=1$), where $i_{n}$ denotes the subband and
$\mathbf{k}_{n}$\ is the in-plane wave vector of the $n$th carrier, with
$n=1,\dots,N$. In such simulations, the physical quantities are typically
computed from the corresponding ensemble averages. The total gain is obtained
from Eq. (\ref{twol_a}) by summing over all carriers $n$ and all available
final states $\left|  j,\mathbf{k}_{n}\right\rangle $ with conserved in-plane
wave vector $\mathbf{k}_{n}$,
\begin{equation}
g\left(  \omega\right)  =\frac{\pi Z_{0}\omega}{Vn_{0}\hbar}\sum_{n,j}%
\frac{\omega_{i_{n}j}}{\left|  \omega_{i_{n}j}\right|  }\left|  d_{i_{n}%
j}\right|  ^{2}\mathcal{L}_{i_{n}j}\left(  \omega\right)  . \label{g}%
\end{equation}
For periodic structures like QCLs, it is appropriate to evaluate transitions
from states within a single central period to the available final states (also
including those in adjacent periods). The simulated volume is then given by
$V=L_{\mathrm{p}}N/n_{\mathrm{s}}$, where $L_{\mathrm{p}}$ and $n_{\mathrm{s}%
}$ are the period length and the sheet doping density per period.

In EMC simulations, all scattering processes are evaluated based on quantum
mechanically calculated Boltzmann-like scattering rates. The contribution from
optically induced transitions can be computed from Eq. (\ref{twol_b}). To
account for effects such as mode competition and multimode lasing, we have to
sum over all relevant longitudinal modes, characterized by their frequencies
$\omega_{m}$ and intensities $I_{m}$. For a carrier sitting in state $\left|
i,\mathbf{k}\right\rangle $, the transition rate to a state $\left|
j,\mathbf{k}\right\rangle $ with occupation probability $f_{j\mathbf{k}}%
$\ thus becomes%
\begin{equation}
r_{i\rightarrow j}=\frac{\pi Z_{0}}{n_{0}\hbar^{2}}\left|  d_{ij}\right|
^{2}\left(  1-f_{j\mathbf{k}}\right)  \sum_{m}I_{m}\mathcal{L}_{ij}\left(
\omega_{m}\right)  .\label{r}%
\end{equation}
In principle, in EMC simulations the photon dynamics can be evaluated in
analogy to the carrier transport, by stochastic sampling of the discrete
photon population in each relevant mode.\cite{2005RPPh...68.2533I} However,
the statistical fluctuations associated with the rare photon emission events
make it difficult to obtain sufficient accuracy.\cite{2005RPPh...68.2533I}
Such problems are here avoided by using classical intensities $I_{m}$ rather
than discrete photon populations.\ The intensity evolution over a time
interval $\delta$ can be described by
\begin{equation}
I_{m}\left(  t+\delta\right)  =I_{m}\left(  t\right)  \exp\left(  \left[
\Gamma\left(  \omega_{m}\right)  g\left(  \omega_{m}\right)  -a\left(
\omega_{m}\right)  \right]  c\delta/n_{0}\right)  ,\label{I}%
\end{equation}
where $c\delta/n_{0}$ is the propagation distance of the light in the gain
medium during that time. In Eq. (\ref{I}), $\delta$\ has to be chosen short
enough that $g\left(  \omega_{m}\right)  $ can be considered constant.
$\Gamma$ denotes the confinement factor, $c$ is the vacuum speed of light, and
$a=a_{\mathrm{m}}+a_{\mathrm{w}}$ is an effective loss coefficient containing
both the mirror and waveguide loss.\cite{2005OExpr..13.3331W} The description
of outcoupling losses by a distributed coefficient $a_{\mathrm{m}}$ is very
common,\cite{2005OExpr..13.3331W} and works especially well for moderate
output coupling at the facets. Interference effects between adjacent modes and
counterpropagating waves are not considered in our approach. Such effects can
become relevant for the multimode dynamics and coherent instabilities in
modelocked lasers.\cite{2007PhRvA..75c1802W}

\begin{figure}[ptb]
\includegraphics{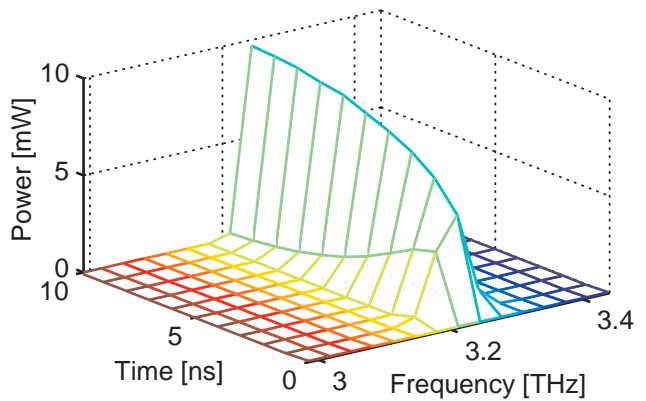}
\caption{(Color online) Temporal evolution of the outcoupled power
$P_{m}^{\mathrm{out}}$ at different longitudinal mode frequencies $f_{m}$.}%
\label{mod}%
\end{figure}

Our self-consistent 3D EMC simulation includes all the essential scattering
mechanisms, also accounting for Pauli's exclusion
principle.\cite{2009JAP...105l3102J} Carrier-carrier interactions are
implemented based on the Born approximation, taking into account screening in
the full random phase approximation.\cite{2005JAP....97d3702B} Also considered
is scattering with impurities as well as acoustic and longitudinal-optical
phonons, including nonequilibrium phonon effects. Interface roughness is
implemented assuming typical values of $0.12\,\mathrm{nm}$ for the mean height
and $10\,\mathrm{nm}$ for the correlation length.\cite{2009JAP...105l3102J} We
simulate four periods of the structure, using periodic boundary conditions for
the first and last period.\cite{2001ApPhL..78.2902I} The optical cavity
dynamics is considered by including the photon-induced scattering
contribution, Eq. (\ref{r}), in the carrier transport simulation and updating
$g\left(  \omega_{m}\right)  $ and $I_{m}$ for each considered mode after
sufficiently short time intervals $\delta$, using Eqs. (\ref{g}) and
(\ref{I}). The homogeneous linewidth in Eq. (\ref{L}) is self-consistently
calculated based on lifetime broadening.\cite{2009JAP...105l3102J} Stimulated
emission and absorption do not affect the optical coherence; the corresponding
scattering events are thus ignored in the linewidth calculation. Inhomogeneous
gain broadening arises if transitions at different frequencies contribute to
the gain spectrum, which can result in multimode operation. The subband
energies and wave functions are obtained by solving the
Schr\"{o}dinger--Poisson (SP) equation.\cite{2009IJQE...45..1059J} The EMC and
SP simulations are then performed iteratively until convergence is reached.

\begin{figure}[ptb]
\includegraphics{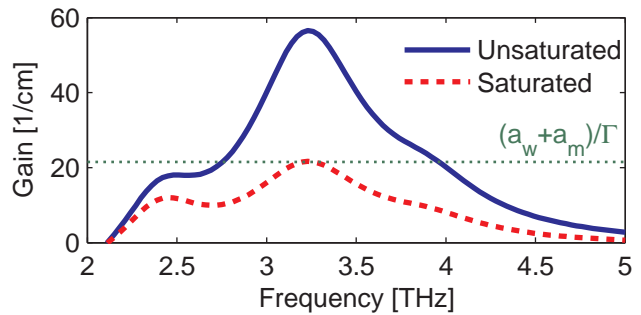}
\caption{(Color online) Unsaturated and saturated power material gain
coefficient vs frequency. For comparison, the cavity loss is also indicated.}%
\label{gain}%
\end{figure}

To validate our simulation approach, we compare the results to experimental
data for a $3.0\,\mathrm{THz}$ resonant phonon depopulation design, consisting
of 178 periods.\cite{2005OExpr..13.3331W} For this structure, comprehensive
temperature resolved experimental data are available, along with detailed
specifications of the device, including the optical cavity. The cavity length
and gain medium cross section are $L=1.22\,\mathrm{mm}$ and $A=178\times
0.0539\,\mathrm{\mu m}\times23\,\mathrm{\mu m}$, respectively, and the facet
reflectivity is $R=0.85$; furthermore, $n_{0}=3.8$, $a_{\mathrm{w}%
}=18.7\,\mathrm{cm}^{-1}$, $a_{\mathrm{m}}=1.3\,\mathrm{cm}^{-1}$, and
$\Gamma=0.93$.\cite{2005OExpr..13.3331W} First, the structure is investigated
for a lattice temperature of $T_{\mathrm{L}}=50\,\mathrm{K}$ at a bias of
$10.7\,\mathrm{kV}$/$\mathrm{cm}$, where our simulation yields the maximum
output power. We consider modes between $2.5\,\mathrm{THz}$ and
$3.5\,\mathrm{THz}$, with a frequency spacing of $df=c/\left(  2Ln_{0}\right)
=0.0324\,\mathrm{THz}$. They are seeded with a small initial intensity
($I_{m}\left(  t=0\right)  =313\,\mathrm{W}$/$\mathrm{cm}^{2}$). For moderate
outcoupling, the single facet power of a mode at frequency $f_{m}=\omega
_{m}/\left(  2\pi\right)  $ is obtained as $P_{m}^{\mathrm{out}}=\frac{1}%
{2}I_{m}A\left(  1-R\right)  /\Gamma$, where the factor $\frac{1}{2}$ arises
because $I_{m}$ encompasses both the forward and backward propagating wave in
the cavity. In Fig. \ref{mod}, the temporal evolution of $P_{m}^{\mathrm{out}%
}$ is shown for 15 of the modes, centered around the gain maximum. After a
time of $10\,\mathrm{ns}$, the lasing operation has nearly reached steady
state, yielding a stationary output power of $10.0\,\mathrm{mW}$ which is
basically accumulated in a single mode. This is in agreement with the
experimentally observed predominant single-mode behavior of this
structure.\cite{2005OExpr..13.3331W} The corresponding experimental value,
measured at a heat sink temperature $T_{\mathrm{s}}=11\,\mathrm{K}%
$\cite{foot1} and uncorrected for the collection efficiency of the Winston
cone used in the measurement setup, is $2.6\,\mathrm{mW}$%
.\cite{2005OExpr..13.3331W} With an estimated collection efficiency of around
30\%,\cite{2005SeScT..20S.222T} this value corresponds to the simulation
result. The unsaturated (solid line) and saturated (dashed line) spectral gain
profiles are shown in Fig. \ref{gain}, and the cavity loss is also indicated
for comparison (dotted line). As expected for stationary lasing, the peak
saturated gain is clamped at the cavity loss.

\begin{figure}[ptb]
\includegraphics{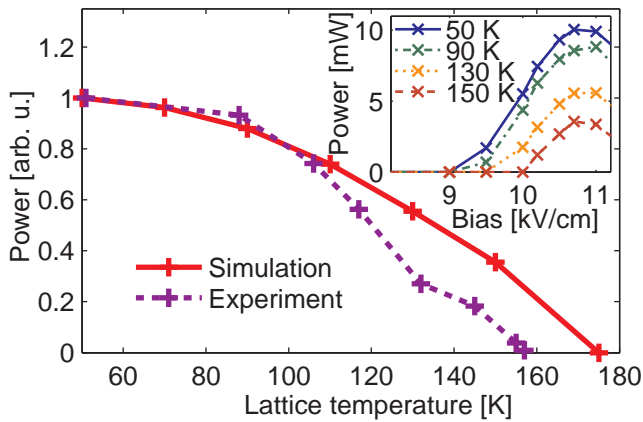}
\caption{(Color online) Simulated and measured maximum outcoupled optical
power vs temperature. The inset shows the simulated optical power vs the
applied bias for different lattice temperatures $T_{\mathrm{L}}$.}%
\label{temp}%
\end{figure}

Multimode simulations, as presented in Fig. \ref{mod}, are computationally
quite tedious, since long simulation times are required to reach steady state.
Thus, in the following, we a priori assume single-mode operation as
experimentally observed,\cite{2005OExpr..13.3331W} considering in our
simulation only the mode with the maximum gain. In Fig. \ref{temp}, the
maximum outcoupled optical power (normalized to its value at $50\,\mathrm{K}$)
is shown as a function of the lattice temperature $T_{\mathrm{L}}$, comparing
simulation results (solid line) to experimental temperature resolved
measurements (dashed line).\cite{2005OExpr..13.3331W,foot1} The overall
agreement is good, with a slight deviation between the simulated and measured
maximum operating temperatures ($175\,\mathrm{K}$ vs $158\,\mathrm{K}$), which
we mainly attribute to uncertainties in the exact value of $a_{\mathrm{w}}%
$.\cite{2005OExpr..13.3331W} The experimental structure lases best at around
$13\,\mathrm{V}$ for all temperatures.\cite{2005OExpr..13.3331W} Also the
simulation yields a temperature insensitive optimum bias which is located at
around $10.7\,\mathrm{kV}$/$\mathrm{cm}$\ (see the inset of Fig. \ref{temp}),
corresponding to a voltage drop of $10.3\,\mathrm{V}$ across the active region
of the investigated structure. This difference between the theoretical and
experimental value can largely be explained by additional parasitic voltage
drops in the experimental structure, especially in the
contacts.\cite{2007JAP...101h6109J} We have also successfully tested our
approach for various other designs, such as a $4.4\,\mathrm{THz}$ high power QCL.\cite{2006ELett..42.89W}

In conclusion, a method is presented to straightforwardly include the optical
cavity field into self-consistent EMC carrier transport simulations, allowing
for the analysis of the actual lasing operation in QCLs. Effects like gain
saturation, gain clamping and mode competition are naturally accounted for.
Comparisons to experimental data confirm the validity of our approach.

C.J. acknowledges support from P. Lugli at TUM. This work was funded by the
Emmy Noether program of the Deutsche Forschungsgemeinschaft (grant
JI115/1-1).

\newpage

\end{document}